\documentclass[journal,twoside]{IEEEtran}
\usepackage{cite}
\usepackage{amsmath,amssymb,amsfonts}
\usepackage{algorithmic}
\usepackage{graphicx}
\usepackage{textcomp}

\usepackage[dvipsnames]{xcolor}
\usepackage{hyperref}
\hypersetup{colorlinks=true,citecolor=ForestGreen,linkcolor=BrickRed,urlcolor=Thistle}
\usepackage{booktabs}
\usepackage{adjustbox}
\usepackage{multirow,rotating}

\def\BibTeX{{\rm B\kern-.05em{\sc i\kern-.025em b}\kern-.08em
    T\kern-.1667em\lower.7ex\hbox{E}\kern-.125emX}}
\begin{document}
\title{Self-Supervised Ultrasound to MRI Fetal Brain Image Synthesis}
\author{Jianbo Jiao, \IEEEmembership{Member, IEEE}, Ana~I.L.~Namburete, Aris~T.~Papageorghiou, and J.~Alison~Noble

\thanks{This work was supported in part by the EPSRC (EP/M013774/1 project Seebibyte, EP/R013853/1 project CALOPUS), the ERC (ERC-ADG-2015 694581, project PULSE), and the NIHR Biomedical Research Centre funding scheme. }
\thanks{Jianbo Jiao, Ana~I.L.~Namburete, and J.~Alison Noble are with the Department
of Engineering Science, University of Oxford, Oxford,
UK (e-mail: jianbo.jiao@eng.ox.ac.uk). }
\thanks{Aris~T.~Papageorghiou is with the Nuffield Department of Women’s \& Reproductive Health, University of Oxford, Oxford, UK.}
}

\maketitle

\begin{abstract}
Fetal brain magnetic resonance imaging (MRI) offers exquisite images of the developing brain but is not suitable for second-trimester anomaly screening, for which ultrasound (US) is employed. Although expert sonographers are adept at reading US images, MR images which closely resemble anatomical images are much easier for non-experts to interpret. Thus in this paper we propose to generate MR-like images directly from clinical US images. 
In medical image analysis such a capability is potentially useful as well, for instance for automatic US-MRI registration and fusion. The proposed model is end-to-end trainable and self-supervised without any external annotations. Specifically, based on an assumption that the US and MRI data share a similar anatomical latent space, we first utilise a network to extract the shared latent features, which are then used for MRI synthesis. Since paired data is unavailable for our study (and rare in practice), pixel-level constraints are infeasible to apply. We instead propose to enforce the distributions to be statistically indistinguishable, by adversarial learning in both the image domain and feature space. To regularise the anatomical structures between US and MRI during synthesis, we further propose an adversarial structural constraint. A new cross-modal attention technique is proposed to utilise non-local spatial information, by encouraging multi-modal knowledge fusion and propagation. We extend the approach to consider the case where 3D auxiliary information (e.g., 3D neighbours and a 3D location index) from volumetric data is also available, and show that this improves image synthesis. The proposed approach is evaluated quantitatively and qualitatively with comparison to real fetal MR images and other approaches to synthesis, demonstrating its feasibility of synthesising realistic MR images.

\end{abstract}


\section{Introduction}
\IEEEPARstart{O}{bstetric} ultrasound (US) is the most commonly applied clinical imaging technique to monitor fetal development. Clinicians use fetal brain US imaging (fetal neurosonography) to detect abnormalities in the fetal brain and growth restriction. However, fetal neurosonography suffers from acoustic shadows and occlusions caused by the fetal skull. On the other hand, magnetic resonance imaging (MRI) is unaffected by the presence of bone and typically provides good and more complete spatial detail of the full anatomy~\cite{pugash2008prenatal}. Whereas MRI is costly and time-consuming, making it unsuitable for fetal anomaly screening, in the second and third trimesters it is often routinely used for assessment of the fetal brain~\cite{bulas2013benefits}.

\begin{figure}[t]
  \centering
  \includegraphics[width=0.72\columnwidth]{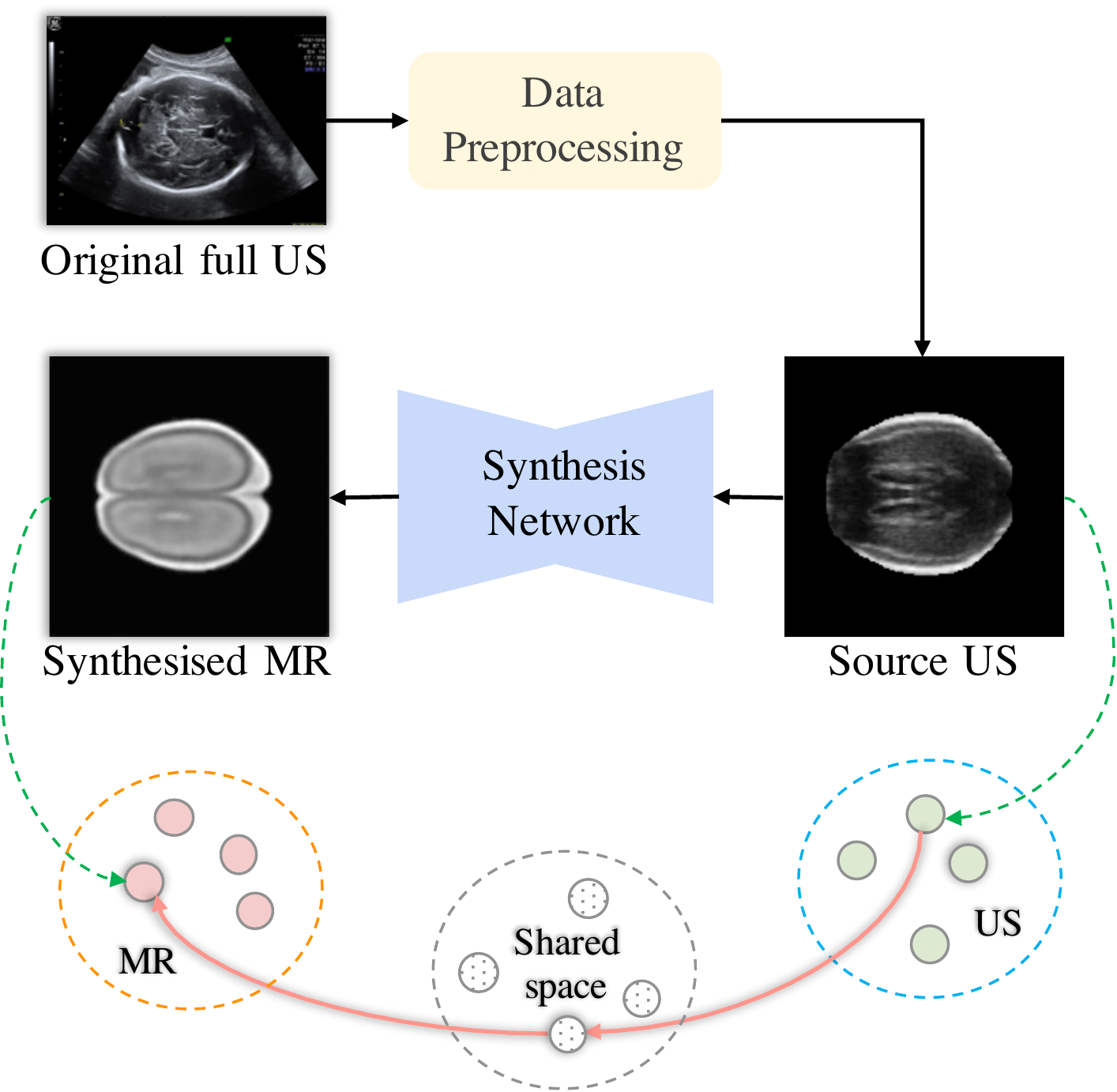}
  \caption{\textbf{Top:} Overview of the proposed US-to-MR synthesis framework. \textbf{Bottom:} Assumption of the shared latent space.}
  \label{fig:pipe}
\end{figure}

Medical image synthesis has received growing interest in recent years.
Most prior work to date has focused on the synthesis between MR and CT (computed tomography) images~\cite{zhao2018towards,nie2017medical,yang2018unpaired} or of retinal images~\cite{costa2018end,costa2017towards}. Simulation of US images has also been proposed to assist in automatic alignment of US and other modalities~\cite{kuklisova2013registration,king2010registering}.
Prior to the deep learning era, medical image synthesis was primarily based on segmentation and atlases. Taking MR-to-CT image synthesis as an example, in segmentation-based methods~\cite{berker2012mri,delpon2016comparison}, first an MR image is segmented into different tissue classes, and then the corresponding synthesised CT image is generated by intensity-filling for each class. On the other hand, atlas-based approaches~\cite{sjolund2015generating,catana2010towards} first register the input MR image to an MR atlas by a transformation, followed by applying such transformation to a CT atlas to synthesise the corresponding CT image. Notwithstanding, the above approaches rely heavily on the segmentation and atlas quality, implying low-quality would directly lead to a poor synthesis. Methods based on convolutional neural networks (CNNs) have demonstrated promising performance for medical image synthesis in recent literature. For instance, given a large number of paired MR-CT data, some proposed methods \cite{zhao2018towards,nie2017medical,roy2017synthesizing} learn a mapping directly from MR to CT through a CNN architecture design. To alleviate the paired data restriction, other methods~\cite{yang2018unpaired,zhang2018translating} have converted the image synthesis problem to image-to-image translation by a recently proposed CycleGAN architecture~\cite{CycleGAN2017}. Even though the training data is not necessarily perfectly registered, weakly paired data (e.g., pairs from the same subject) or other types of supervision from additional tasks like dense segmentation are still required in these methods.

In this paper, we address the problem of US-to-MR image synthesis by a learning-based framework. Fig.~\ref{fig:pipe} summarises the proposed framework. We design an anatomically constrained self-supervised model to learn the mapping from US to MR under an assumption that US and MR share a common representation in a latent space.
The anatomical constraint considers both the latent space and the geometrical structures between the two modalities. To the best of our knowledge, this article presents the first attempt towards unpaired US-to-MR synthesis in a self-supervised manner. Qualitative and quantitative experiments demonstrate that the proposed approach generates realistic MR images, even with highly-imbalanced data. 

\emph{{Relationship to Preliminary Work~\cite{Jianbo2019anatomy}}:}
An early version of this work was presented in~\cite{Jianbo2019anatomy}. In the current paper, we considerably expand the preliminary study by: 1) We further propose three new solutions to utilise 3D auxiliary information and boost the synthesis performance. Leveraging additional neighbouring inputs and predicting the position in 3D space as an auxiliary task are explored to achieve the goal. 2) In this version we provide a more detailed analysis of our framework and its new extensions. The detailed architecture of each network components are elaborated; more details with additional illustrations are included for the EdgeNet; new perspectives on the proposed cross-modal attention is included; a detailed analysis of the anatomical latent space is presented. 3) Additional experimental evaluations are included in this extension, with more training details; standard deviation for all the quantitative performance is reported for better understanding of the models; more qualitative results are presented with comparison to other solutions; an anatomy-preserving analysis is presented to evaluate the effectiveness of the proposed approach for both synthetic structures and real data; performance on the aforementioned 3D-based solutions is also reported with comparison to our preliminary results.

The main highlights of the paper are summarised as:
\begin{itemize}
  \item
  We present an approach to synthesise MR-like images from unpaired US images;
  \item
  We propose an anatomy-aware deep neural network architecture with mono-directional consistency, to address the synthesis problem in a self-supervised manner;
  \item
  Based on the shared latent space assumption, we propose a latent space consistency constraint;
  \item
  We propose a cross-modal attention module that propagates information across modalities in the feature domain;
  \item
  We propose to leverage 3D auxiliary information to reduce ambiguity during image synthesis;
  \item
  Comprehensive experimental evaluation and analysis show that the proposed synthesis framework is able to generate high-quality MR-like images and performs favourably against other alternative methods.
\end{itemize}

The rest of this paper is organised as follows. Section~\ref{sec:related} reviews related work and discusses the main differences to our approach. In Section~\ref{sec:method}, we elaborate on the detail of our self-supervised US-to-MR image synthesis approach, with analysis of the network architecture design and the underlying representative features. Following that, we perform extensive experiments evaluating the effectiveness of the proposed framework both qualitatively and quantitatively in Section~\ref{sec:exp}. In addition, potential applications derived from this work and the model generalisations are discussed in Section~\ref{sec:app}. Finally, the paper is concluded in Section~\ref{sec:conc} and possible future directions are discussed.

\section{Related Work}\label{sec:related}
\subsection{Medical Image Synthesis}
Medical image synthesis or simulation aims to synthetically generate one imaging modality from another. 
Classical methods to achieve this have been based on segmentation and atlases. Segmentation-based approaches are straightforward, and in the case of CT synthesis from MRI, may, for instance, first segment the MR images into different parts (e.g., bony structure, soft tissue) and then assign the corresponding CT number to each part. In~\cite{lee2003radiotherapy}, the authors study the possibility of radiotherapy treatment planning using only MR by bone and water segmentation. Berker et al. propose to address the MRI-based attenuation correction problem by segmenting air, bone and tissues~\cite{berker2012mri}.  On the other hand, atlas-based methods generate a synthetic CT by deforming a CT atlas onto the patient space, where the required deformation is found by registering an MR atlas to the patient real MR image. Dowling et al.~\cite{dowling2012atlas} generate pseudo-CT scans by nonrigid registration of an MRI atlas to an MRI scan. The authors of~\cite{sjolund2015generating} use atlas-based regression to deform a collection of atlas CTs into a single pseudo-CT, based on the target MR and an atlas database.
However, the segmentation-based approaches suffer from requiring a time-consuming segmentation, while the uncertainty in registration (e.g., missing tissues) is a critical inherent limitation of atlas-based methods. With the recent progress of deep learning in medical image analysis, CNN-based approaches have started to dominant the medical image synthesis. Zhao et al.~\cite{zhao2018towards} propose to directly optimise the mapping function for MR to CT synthesis, with reference to different 3D views. However, such a regression-based method may lead to blurred results if there are misalignments between CT and MR images. To handle this problem, some works~\cite{nie2017medical,costa2018end} augment the regression-based loss with another adversarial loss in a generative adversarial network framework~\cite{goodfellow2014generative}. Although blur caused by misalignment has been addressed, a training set of paired images (e.g., MRI and CT) is still necessary for the above models. Whereas such paired training data is very scarce for many clinical imaging applications.

\subsection{Self-Supervised Learning}
Self-supervised (also termed `unsupervised' in some literature) learning is a learning technique that does not rely on the supervision from external label/annotation, i.e., the learning is totally based on the available data itself. An auto-encoder (AE)~\cite{bengio2007greedy} is one of the most basic self-supervised learning approaches, which optimises a self-reconstruction loss by recovering the input signal itself. Derivatives including the denoising auto-encoder (DAE)~\cite{vincent2010stacked} and variational auto-encoder (VAE)~\cite{kingma2013auto} also focus on self-supervised learning but with different optimisation functions. Recently, Goodfellow et al. proposed the Generative Adversarial Network (GAN)~\cite{goodfellow2014generative} which learns to generate meaningful signals from random noise, by playing a minimax game. Based on the GAN framework, an architecture named CycleGAN ~\cite{CycleGAN2017} or DualGAN~\cite{yi2017dualgan} is proposed to address the problem of image-to-image translation, without the dependency on paired training data. Consequently, such an architecture with cycle consistency {has enabled} a group of medical image synthesis methods in recent years, mainly focusing on MR-to-CT image synthesis~\cite{yang2018unpaired,zhang2018translating,wolterink2017deep,hiasa2018cross}, and vice versa~\cite{zhang2018translating,chartsias2017adversarial}. Building on top of the original CycleGAN, in~\cite{yang2018unpaired} and \cite{hiasa2018cross}, additional loss terms have been introduced to further constrain the structure features. Zeng and Zheng~\cite{zeng2019hybrid} propose a hybrid GAN that combines a 3D generator and a 2D discriminator to synthesise CT from MR images, in a weakly-supervised manner. In~\cite{zhang2018translating}, the authors propose a solution to MR-CT synthesis by a 3D CNN composed of mutually beneficial generators and segmentors with cycle- and shape-consistency. Paired data dependency has been alleviated to some extent by the above CycleGAN-based approaches. However, aligned or weakly-aligned data or auxiliary tasks are still necessary in these works. Besides, MR and CT are relatively, similar in anatomically appearance and relatively easier to align, when compared with ultrasound. To our knowledge, there is no prior work on cross-modal image synthesis from ultrasound (US) data, in a data-driven self-supervised manner.

\subsection{Ultrasound Image Analysis}
Different from the aforementioned medical imaging modalities MRI and CT, US imaging usually does not present as clear and sharp anatomical structures. However, its real-time and un-harmful properties make it a much more suitable choice for many medical screening scenarios, including fetal development monitoring. 
Prior fetal US image analysis work mainly focuses on fetal anatomy detection~\cite{maraci2014searching,yaqub2015guided,chen2015standard,baumgartner2017sononet,cai2018multi} and registration to other modalities~\cite{kuklisova2013registration,king2010registering,wein2008automatic,xiao2019evaluation}. Maraci et al.~\cite{maraci2014searching} propose an approach to make the US diagnosis easier by combining simple US scanning protocols with machine learning solutions. Yaqub et al.~\cite{yaqub2015guided} propose a random forest based classifier to categorise fetal US images. With the help of deep learning techniques, Chen et al.~\cite{chen2015standard} present a CNN-based approach to locate the fetal abdominal standard plane in US videos. Some methods~\cite{cai2018multi} utilise human eye-gaze data to assist standard plane detection. The fusion of tracked US with other modalities like CT and MRI has benefits for a variety of clinical applications. Wein et al.~\cite{wein2008automatic} develop methods to simulate US from CT in real-time, while in~\cite{xiao2019evaluation} the authors evaluate the performance of methods of MRI to US registration. Kuklisova et al.~\cite{kuklisova2013registration} propose a method for 3D fetal brain US and MRI registration by simulating a pseudo-US from an MR volume segmentation. While most existing work focusing on the above US image analysis topics, there lacks a study on US to MRI synthesis in the literature.

\begin{figure*}[t]
  \centering
  \includegraphics[width=0.96\textwidth]{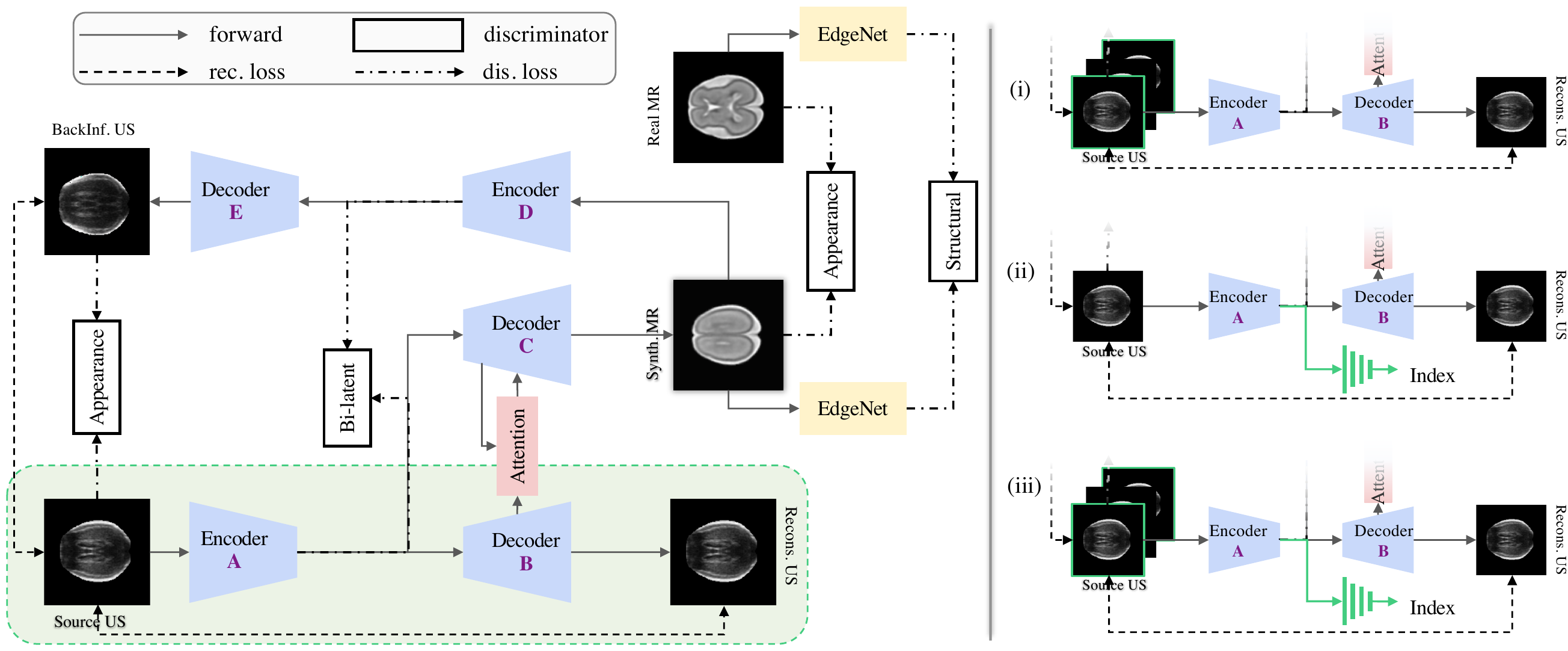}
  \caption{\textbf{Left:} Architecture of the proposed US-to-MR synthesis framework. \textbf{Right:} Solutions to leverage 3D auxiliary information, each of which can be plugged into the green part in the left framework.(i) Augmented with neighbouring slices in the 3D volume; (ii) Predicting the index of the slice in the 3D volume; (iii) Both with augmented slices and index prediction.}
  \label{fig:frame}
\end{figure*}

\section{Method}\label{sec:method}
In this section, we elaborate the proposed approach for US-to-MR image synthesis in detail. Specifically, we first pre-process the US volumes by cropping and automatically aligning the US volumes as described in~\cite{namburete2018fully}. Following that, we manually align the MR volume to the same reference space. Then we propose a novel learning-based framework for unpaired US-to-MR synthesis, which is illustrated in Fig.~\ref{fig:frame} (detailed structure of the blue block in Fig.~\ref{fig:pipe}(a)). Given a source US image, the corresponding MR image is synthesised with reference to real MR data.
In addition to pixel-level (\emph{rec. loss}) constraints, a distribution similarity (\emph{dis. loss}) is also incorporated to address the unpaired data property and ensure anatomical consistency.
As our objective is to synthesis MR from US images, the overall design of the proposed framework is mono-directional, instead of bi-directional as in the CycleGAN architecture~\cite{CycleGAN2017}. That is, we only have the forward cycle (i.e., US$\rightarrow$MR$\rightarrow$US) without the reverse cycle (i.e., MR$\rightarrow$US$\rightarrow$MR), and we experimentally find that such a design leads to less ambiguity for image synthesis in our case.
Next, we elaborate on each of the proposed components in detail.

\subsection{Anatomy-Aware Synthesis}
Paired (e.g., same fetus at the same gestational age) fetal brain US and MR data is rare in clinical practice, and unavailable in our case. Even if it is available, US and MR are not simultaneously co-registered as has often been assumed in prior medical image synthesis methods~\cite{zhao2018towards,nie2017medical}. Hence it is infeasible to learn the mapping from US to MR directly by traditional CNN-based techniques for our task. Therefore, we propose to address the problem through a synthesis framework, by enforcing the synthesised MR images to lie in a similar distribution to real MR data. Throughout the synthesis, an important objective is to correctly map the clinically important anatomical structures between the two modalities. As a result, anatomy-aware constraints are specifically designed to implicitly preserve anatomy consistency.

\subsubsection{Anatomical Feature Extraction}
As paired data is unavailable, we assume that the US and MR images share an anatomical latent space (Fig.~\ref{fig:pipe}(b)). Building upon this assumption, instead of using the pixels in the image domain, we propose to extract the underlying anatomical features and synthesise images in the corresponding MR domain accordingly. Specifically, we leverage an autoencoder to extract the latent features, as shown in the bottom-left part of Fig.~\ref{fig:frame} (encoder-A$\rightarrow$decoder-B). Assume the set of $n$ source US images as $\left\{x_U^i\right\}_{i=1}^n$ where $x_U^i \in \mathcal{X}_U$ is the $i^{th}$ image, the extracted anatomical feature is formally defined as $y^i=F(x_U^i)$ where $F(\cdot)$ is the encoder.

\subsubsection{Bi-directional Latent Space Consistency}
The above extracted latent features are fed into decoder-C to synthesise the corresponding MR image. As pixel-level supervision is unavailable for \textit{Synth. MR}, we use a backward-inference path (encoder-D$\rightarrow$decoder-E) to recover the source US. Denoting the encoded latent feature (at the end of encoder-D) as $y_b^i$, we propose a bi-directional latent space consistency constraint, based on the assumption of shared latent space. As a result, $y^i$ and $y_b^i$ are forced to lie in a similar distribution by means of adversarial learning (\textit{Bi-latent} block in Fig.~\ref{fig:frame}).

\subsubsection{Structural Consistency}
Although the anatomical feature extraction module encodes the main structure of an US image, the image structure in the MR domain is quite different in appearance compared to that in the US domain. To synthesise realistic MR images, we further propose a constraint to enforce the structures of the \textit{Synth. MR} and the \textit{Real MR} to be similar. Noting the unpaired nature of our data, we choose to constrain the structural information to lie in a similar distribution. Specifically, the edge information of the \textit{Synth. MR} and \textit{Real MR} is extracted by an EdgeNet, following which  a structural discriminator (\textit{Structural} block in Fig.~\ref{fig:frame}) is leveraged to compute the edge similarity.

\subsection{Cross-modal Attention}
\begin{figure}[t]
  \centering
  \includegraphics[width=0.9\columnwidth]{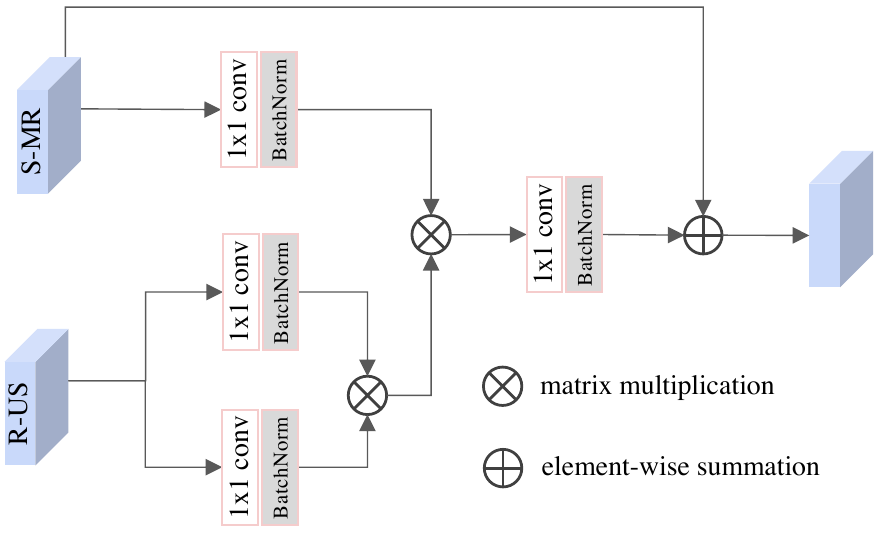}
  \caption{Detailed architecture of the proposed cross-modal attention module.}
  \label{fig:att}
\end{figure}

Based on the aforementioned components, the MR image synthesis process is mainly guided by the latent features $y^i$ extracted from encoder-A. To further leverage guidance across different modalities, we propose a cross-modal attention module between the US decoder-B and the MR decoder-C, as shown in Fig.~\ref{fig:frame} (the red \emph{Attention} block). Specifically, the US features are reformulated as self-attention guidance for MR image synthesis, and such a guidance is applied to the MR features implicitly in an attentive manner (detailed in Fig.~\ref{fig:att}). The cross-modal attention module consists of several $1\times1$ convolutional layers and a skip connection, without any modification to the input feature dimension.
This cross-modal attention module leverages guidance across different modalities (MR and US here) by cross-referencing the features from the two modalities. Specifically, the features from US are combined with the features from MR by matrix multiplication, which can also be considered as an approximation of the mutual information acquisition across these two modalities.
The $1\times1$ convolutions adapt the channel dimension for the subsequent multiplication. Since our target is to synthesise MR-like images, the original MR features are added back to the mutual information by a skip connection, which also keeps the feature dimension.
A similar idea for single modality attention (also termed as self-attention~\cite{zhang2018self} or non-local scheme~\cite{wang2018non}) has been shown to be effective to leverage neighbouring information in {natural image analysis}.

Denoting the features from the US reconstruction decoder-B (R-US in Fig.~\ref{fig:att}) as $f_U$ and the features from the MR synthesis decoder-C (S-MR in Fig.~\ref{fig:att}) as $f_M$, the updated feature after the cross-modal attention module is defined as:
\begin{equation}
  \widetilde{f}_M=\eta(\delta(f_U)^T \otimes \phi(f_U) \otimes g(f_M))+f_M,
\end{equation}
where $\eta,\delta,\phi,g$ are linear embedding functions which are implemented by $1\times1$~convolutions. The proposed cross-modal attention enforces the network to not only focus on local activations (favoured by CNNs) but also non-local context information via both self- and cross-modal attention.
Specifically, the local and non-local correlation is modelled by feature matrix multiplication ($\otimes$).

\subsection{Joint Adversarial Objective Function}
In this section, we now formally define the objective function that is used to train the proposed US-to-MR synthesis model. As aforementioned (and can be observed in Fig.~\ref{fig:frame}), there are two forms of objective terms: one based on pixel-wise reconstruction (\emph{rec. loss}) and the second a discriminator-based distribution similarity (\emph{dis. loss}). The \emph{rec. loss} is defined as an $\ell1$-norm, while the \emph{dis. loss} is achieved by a discriminator with adversarial learning~\cite{goodfellow2014generative}.

\subsubsection{Generative Adversarial Learning}
The generative adversarial network (GAN) is a self-supervised learning framework proposed in~\cite{goodfellow2014generative}, which consists of two modules, namely a generator and a discriminator. The main idea of a GAN is to play a minimax game with these two modules. The training of a GAN forces the distribution of the source data $x$ to be similar to the target data ($y$) distribution. Suppose the mapping from $x$ to a new data space is $G(x;\theta_g)$ where $G$ is the generator with parameters $\theta_g$. The discriminator $D(y;\theta_d)$ outputs a binary scalar indicating whether the data $y$ is real or fake. Then the $D$ net is trained to maximise the probability of assigning the correct label to both real data samples and samples from $G$. Meanwhile, the $G$ net is trained to minimise $\log(1-D(G(x)))$. The final objective function for a GAN is defined as: $\min _{G} \max  _{D}\mathbb{E}[\log D(y)]+\mathbb{E}[\log(1-D(G(x)))]$.

\subsubsection{Proposed Joint Objective Function}
Denoting the reconstructed US as $\hat{x}_U \in \hat{\mathcal{X}_U}$, latent feature as $y \in \mathcal{Y}$, synthesised MR as $\hat{x}_M \in \hat{\mathcal{X}}_M$, and real MR as $x_M \in \mathcal{X}_M$. The objective for the forward path (US$\rightarrow$MR) is defined as:
\begin{equation}\label{eq:2}
\{\min\mathcal{L}_F~|~\mathcal{L}_F=\lambda\mathcal{L}_{lat}+\gamma\mathcal{L}_{app}+\gamma\mathcal{L}_{stru}\},
\end{equation}
where
\begin{equation}
  \mathcal{L}_{lat}=\mathbb{E}_{x_U\in\mathcal{X}_U}\|G_U(F(x_U))-x_U\|_1,
\end{equation}
\begin{equation}
  \begin{split}
    \mathcal{L}_{app}&=\mathbb{E}_{x_M\in\mathcal{X}_M}(D_{app}(x_M))\\
    &+\mathbb{E}_{y\in\mathcal{Y}}\log(1-D_{app}(G_M(y))),
  \end{split}
\end{equation}
\begin{equation}
  \begin{split}
  \mathcal{L}_{stru}&=\mathbb{E}_{x_M\in\mathcal{X}_M}(D_{stru}(E(x_M)))\\
  &+\mathbb{E}_{y\in\mathcal{Y}}\log(1-D_{stru}(E(G_M(y)))).
  \end{split}
\end{equation}

Here the $\mathcal{L}_{lat},\mathcal{L}_{app},\mathcal{L}_{stru}$ are loss terms for the reconstruction from latent space, appearance, and structural consistency, respectively. The first term $\mathcal{L}_{lat}$ represents the generator loss while the following two terms indicate the discriminator loss.
The decoder-B that is used to reconstruct \emph{Recons.US} is represented by $G_U$, while the decoder-C for MR synthesis is represented by $G_M$. The $\hat{x}_U$ and $\hat{x}_M$ are defined as: $\hat{x}_U=G_U(F(x_U))$, $\hat{x}_M=G_M(y)$. The discriminators $D_{app}$ and $D_{stru}$ that each consists of four \textit{conv} layers are used to measure the similarity for appearance and structure, respectively. The EdgeNet is represented by $E$, while parameters $\lambda$ and $\gamma$ are balancing weights for the objective terms, so that they lie on a similar scale.

\begin{table*}[]
  \centering
  \caption{{Details of the network design. The five main components (subnetworks) are presented with detailed parameter settings. The numbers follow each \textit{conv/up-conv} are the kernel size and number of channels, while the number follow \textit{maxpool} is the scale. \emph{FC} represents {a} fully-connected layer and the numbers following are the input and output dimensions. The symbol $\odot$ represents concatenation, while $\oplus$ and $\otimes$ the element-wise summation and matrix multiplication.}}
  \label{tab:net}
  \begin{adjustbox}{max width=\textwidth}
    \begin{tabular}{@{}lll|lll|lll@{}}
  \toprule
  \multicolumn{3}{c|}{Encoder}                                                                                                   & \multicolumn{3}{c|}{Cross-modal Attention}                                                                             & \multicolumn{3}{c}{Discriminator}                                                                 \\ \midrule
  Layer  & Input            & Parameter                                                                                          & Layer                & Input                                                                          & Parameter      & Layer  & Input                                                          & Parameter                \\ \midrule
  enc\_1  & US/MR img        & \begin{tabular}[c]{@{}l@{}}conv, 3$\times$3, 48\\ conv, 3$\times$3, 48\\ maxpool, 2\end{tabular}                 & convMR             & S-MR                                                                           & conv, 1$\times$1, 72  & disc\_1 & \begin{tabular}[c]{@{}l@{}}img/\\ edge/\\ feature\end{tabular} & conv, 3$\times$3, 64, stride=2  \\
  enc\_2  & enc\_1            & \begin{tabular}[c]{@{}l@{}}conv, 3$\times$3, 48\\ maxpool, 2\end{tabular}                                 & convUS\_1          & R-US                                                                           & conv, 1$\times$1, 72  & disc\_2 & disc\_1                                                         & conv, 3$\times$3, 128, stride=2 \\
  enc\_3  & enc\_2            & \begin{tabular}[c]{@{}l@{}}conv, 3$\times$3, 48\\ maxpool, 2\end{tabular}                                 & convUS\_2          & R-US                                                                           & conv, 1$\times$1, 72  & disc\_3 & disc\_2                                                         & conv, 3$\times$3, 256, stride=2 \\
  enc\_4  & enc\_3            & \begin{tabular}[c]{@{}l@{}}conv, 3$\times$3, 48\\ maxpool, 2\end{tabular}                                 & fuse                 & \begin{tabular}[c]{@{}l@{}}convUS\_1$\otimes$\\ convUS\_2$\otimes$\\ convMR\end{tabular} & conv, 1$\times$1, 144 & disc\_4 & disc\_3                                                         & conv, 3$\times$3, 512           \\
  enc\_5  & enc\_4            & \begin{tabular}[c]{@{}l@{}}conv, 3$\times$3, 48\\ maxpool, 2\end{tabular}                                 & output & S-MR$\oplus$fuse & - & out    & disc\_4                                                         & conv, 3$\times$3, 1             \\ \midrule
  \multicolumn{3}{c|}{Decoder}                                                                                                   & \multicolumn{3}{c|}{EdgeNet}                                                                                           & \multicolumn{3}{c}{Volume Index Predictor}                      \\ \midrule
  Layer  & Input            & Parameter                                                                                          & Layer                & Input                                                                          & Parameter      & Layer       & Input                                                               & Parameter                         \\ \midrule
  dec\_1  & enc\_5            & \begin{tabular}[c]{@{}l@{}}conv, 3$\times$3, 48\\ up-conv, 3$\times$3, 48, stride=2\end{tabular}                 & g\_hori              & \begin{tabular}[c]{@{}l@{}}synthetic MR/\\ real MR \end{tabular}                                                             & conv, 1$\times$5, 1   & Ind\_1       & enc\_5                                                               & conv, 3$\times$3, 48                         \\
  dec\_2  & dec\_1$\odot$enc\_4 & \begin{tabular}[c]{@{}l@{}}conv, 3$\times$3, 96\\ conv, 3$\times$3, 96\\ up-conv, 3$\times$3, 96, stride=2\end{tabular} & g\_vert              & g\_hori                                                                        & conv, 5$\times$1, 1   & Ind\_2       & Ind\_1                                                               & conv, 3$\times$3, 32                         \\
  dec\_3  & dec\_2$\odot$enc\_3 & \begin{tabular}[c]{@{}l@{}}conv, 3$\times$3, 96\\ conv, 3$\times$3, 96\\ up-conv, 3$\times$3, 96, stride=2\end{tabular} & s\_hori              & g\_vert                                                                        & conv, 3$\times$3, 1   & Ind\_3       & Ind\_2                                                               & conv, 3$\times$3, 16                         \\
  dec\_4  & dec\_3$\odot$enc\_2 & \begin{tabular}[c]{@{}l@{}}conv, 3$\times$3, 96\\ conv, 3$\times$3, 96\\ up-conv, 3$\times$3, 96, stride=2\end{tabular} & s\_vert              & s\_hori                                                                        & conv, 3$\times$3, 1   & Ind\_4       & Ind\_3                                                               & conv, 3$\times$3, 4                         \\
  dec\_5  & dec\_4$\odot$enc\_1 & \begin{tabular}[c]{@{}l@{}}conv, 3$\times$3, 96\\ conv, 3$\times$3, 96\\ up-conv, 3$\times$3, 96, stride=2\end{tabular} &                      &                                                                                &                & Out\_Ind       & Ind\_4                                                               & FC, 16, 2                         \\
  dec\_6  & dec\_5$\odot$US   & \begin{tabular}[c]{@{}l@{}}conv, 3$\times$3, 64\\ conv, 3$\times$3, 32\end{tabular}                              &                      &                                                                                &                &        &                                                                &                          \\
  output & dec\_6            & conv, 3$\times$3, out\_channel                                                                            &                      &                                                                                &                &        &                                                                &                          \\ \bottomrule
  \end{tabular}
\end{adjustbox}
\end{table*}

Denoting the \emph{BackInf.US} recovered from the \textit{Synth.MR} as $\widetilde{x}_U \in \widetilde{\mathcal{X}}_U$ and the back-inferred feature at the end of encoder-D as $y^{back}\in\mathcal{Y}^{back}$, the objective for the backward path (Synth.MR$\rightarrow$BackInf.US) is defined as:
\begin{equation}
  \{\min\mathcal{L}_B~|~\mathcal{L}_B=\lambda\mathcal{L}_{proj}+\gamma\mathcal{L}_{app}^{back}+\gamma\mathcal{L}_{bi}\},
\end{equation}
where
\begin{equation}
  \mathcal{L}_{proj}=\mathbb{E}_{\tilde{x}_U\in\tilde{\mathcal{X}}_U,x_U\in\mathcal{X}_U}\|\tilde{x}_U-x_U\|_1,
\end{equation}
\begin{equation}
  \begin{split}
    \mathcal{L}_{app}^{back}&=\mathbb{E}_{x_U\in\mathcal{X}_U}(D_{app}^{back}(x_U))\\
    &+\mathbb{E}_{y^{back}\in\mathcal{Y}^{back}}\log(1-D_{app}^{back}(G_{BU}(y^{back}))),
  \end{split}
\end{equation}
\begin{equation}
  \mathcal{L}_{bi}=\mathbb{E}_{y\in\mathcal{Y}}(D_{bi}(y))+\mathbb{E}_{y^{back}\in\mathcal{Y}^{back}}\log(1-D_{bi}(y^{back})).
\end{equation}

Here the $\mathcal{L}_{proj},\mathcal{L}_{app}^{back},\mathcal{L}_{bi}$ are loss terms for the back-inference reconstruction, backward appearance, and bi-directional latent space consistency, respectively. Similar to Eq.~\ref{eq:2}, parameters $\lambda$ and $\gamma$ are balancing weights.
The decoder-E that is used to back recover \emph{BackInf.US} is represented by $G_{BU}$ and $\tilde{x}_U=G_{BU}(y^{back})$. The discriminators $D_{app}^{back}$ and $D_{bi}$ are used to compute the similarity for backward-inference and bi-directional latent space, respectively.
Based on the above defined objective terms, the final joint loss function for our model training is defined as:
\begin{equation}
  \mathcal{L}=\mathcal{L}_F+\mathcal{L}_B.
\end{equation}

\subsection{3D Auxiliary Information}\label{sec:3D}
Here we investigate the possibility of leveraging 3D volumetric information to improve the synthesis. The proposed approaches to leverage 3D information are shown in Fig.~\ref{fig:frame}-Right. Specifically, we propose three strategies to achieve the goal: 1) by adding neighbouring slices as augmented input; 2) by predicting the position/index of the current slice in the volume as an additional task; 3) with both the augmented input and the index prediction task. For simplicity, we only show the modified part (the green block in Fig.~\ref{fig:frame}-Left) compared to the 2D-based model. These approaches are motivated by the constraints humans appear to have when viewing slice-wise volumetric data. We assume that if the model is able to utilise the 3D positional information (by either referring to neighbours or directly reasoning its position), it could have a more thorough understanding of the whole anatomical structure and alleviate synthesis ambiguity. The above modifications do not severely influence the original network architecture. The augmented input only leads to a channel number update ($1\rightarrow3$) for the first layer of Encoder-A, while all the decoders only output the middle slice. The index prediction branch is implemented by four convolutional layers with a fully-connected layer, in a regression manner. In the rest of this paper, we use the original 2D settings (by default) as aforementioned, unless otherwise specified.

\subsection{Network Architecture}
The detailed network architecture design and parameters are presented in Table~\ref{tab:net}. The proposed network basically consists of convolutional (\textit{conv}) layers, up-convolutional (\textit{up-conv}) layers, and pooling (\textit{maxpool}) layers. Specifically, each part is described as follows:

\subsubsection{Encoder and Decoder}
All the encoders (and the decoders) share the same architecture as shown in the left part of Table~\ref{tab:net}. The encoder takes either the US image or MR image as input and consists of five \emph{enc} blocks. On the other hand, the decoder is composed of a mirrored architecture to the encoder to reconstruct/synthesise the target image. Each \emph{enc} block consists of \emph{conv} layers and \emph{maxpool} layers, while the \emph{dec} block in the decoder mainly consists of \emph{conv} layers and \emph{up-conv} layers. Skip connections are added between the encoder and decoder.

\subsubsection{Cross-modal Attention}
The cross-modal attention module is proposed to implicitly learn the feature fusion strategy between the US features and MR features. As illustrated in Fig.~\ref{fig:att} and Table~\ref{tab:net}, this module is implemented by several basic $1\times1$ \emph{conv} layers, with matrix multiplication. The \emph{conv} layers are utilised to adjust the feature dimension and combination weights for the following fusion.

\begin{figure}[t]
  \centering
  \includegraphics[width=0.9\columnwidth]{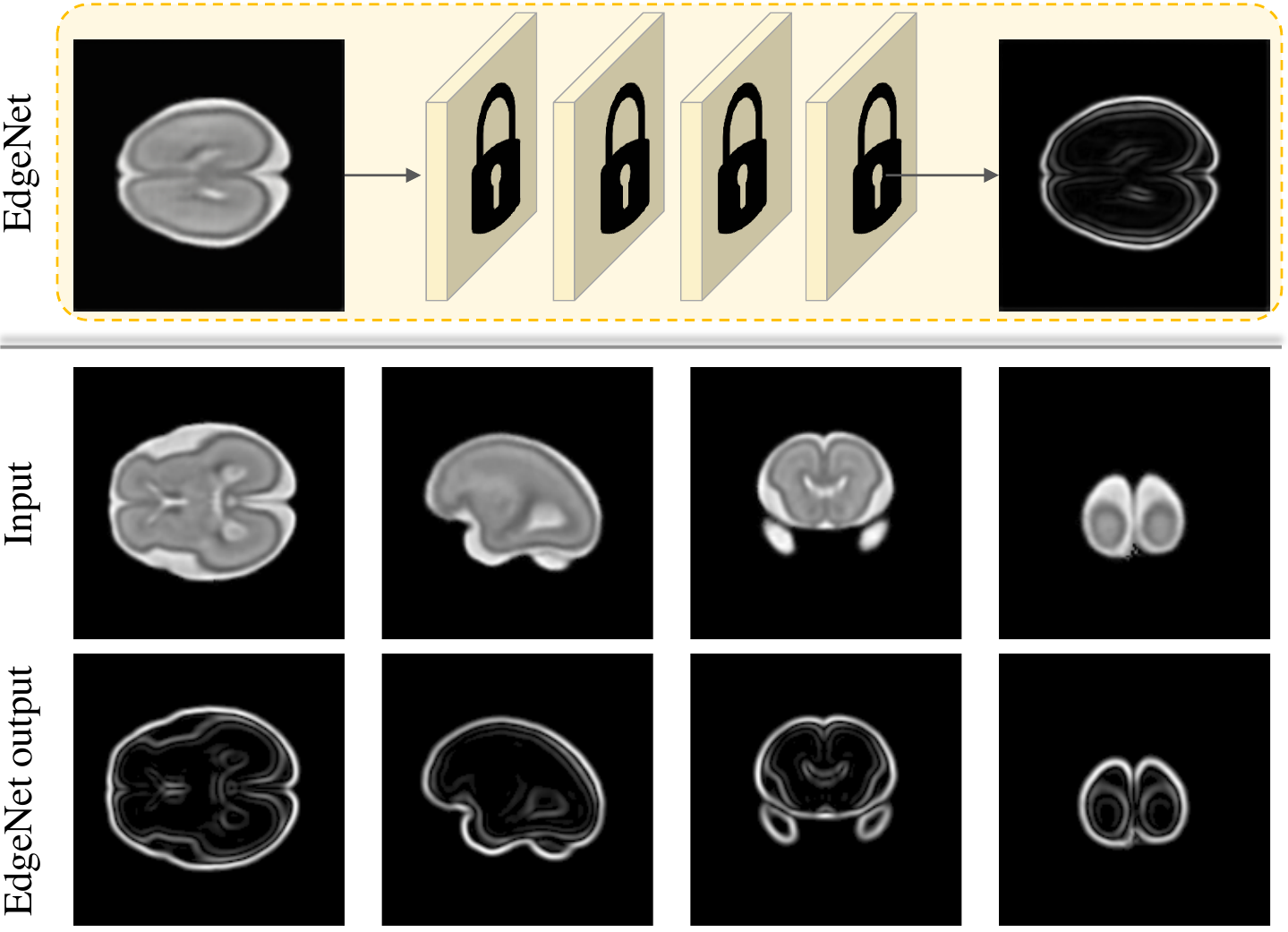}
  \caption{\textbf{Top:} Illustration of the EdgeNet, where the lock symbol indicates {a} layer frozen. \textbf{Bottom:} Example results, where the first row shows the input MR images and the second row shows the detected edges from the EdgeNet.}
  \label{fig:edgenet}
\end{figure}

\subsubsection{EdgeNet}
The EdgeNet extracts edges from the input image (either US or MRI). It consists of four \emph{conv} layers, imitating the Canny edge detector (can also be other edge detectors). Specifically, the input images are first smoothed with Gaussian kernels and convolved with Sobel edge filters, followed by non-maximum suppression and thresholding. Since the edge maps extracted from the EdgeNet are further fed into the discriminator for loss computation, all the parameters in this module are fixed, avoid updating.
Illustration of the proposed EdgeNet, together with example generated edge maps are shown in Fig.~\ref{fig:edgenet} for reference.

\subsubsection{Discriminator}
There are four discriminators in our main architecture, measuring the similarity in terms of appearance, latent space, and structure. Each discriminator is composed of five \emph{conv} layers and outputs a scalar value, indicating whether the input is real or fake. The discriminator essentially acts as a binary classifier.

\subsubsection{Volume Index Predictor}
The index prediction branch consists of four \emph{conv} layers and an FC layer. All the \emph{conv} layers are with {a} $3\times3$ kernel size and decrease in the channel dimension till the final fully-connected output layer. The index predictor takes the feature at the latent space as input and predicts the corresponding index of the original input US. Possible solutions of leveraging the index predictor are illustrated in Fig.~\ref{fig:frame}-Right.

\subsection{Anatomical Space Analysis}
\begin{figure}[t]
  \centering
  \includegraphics[width=\columnwidth]{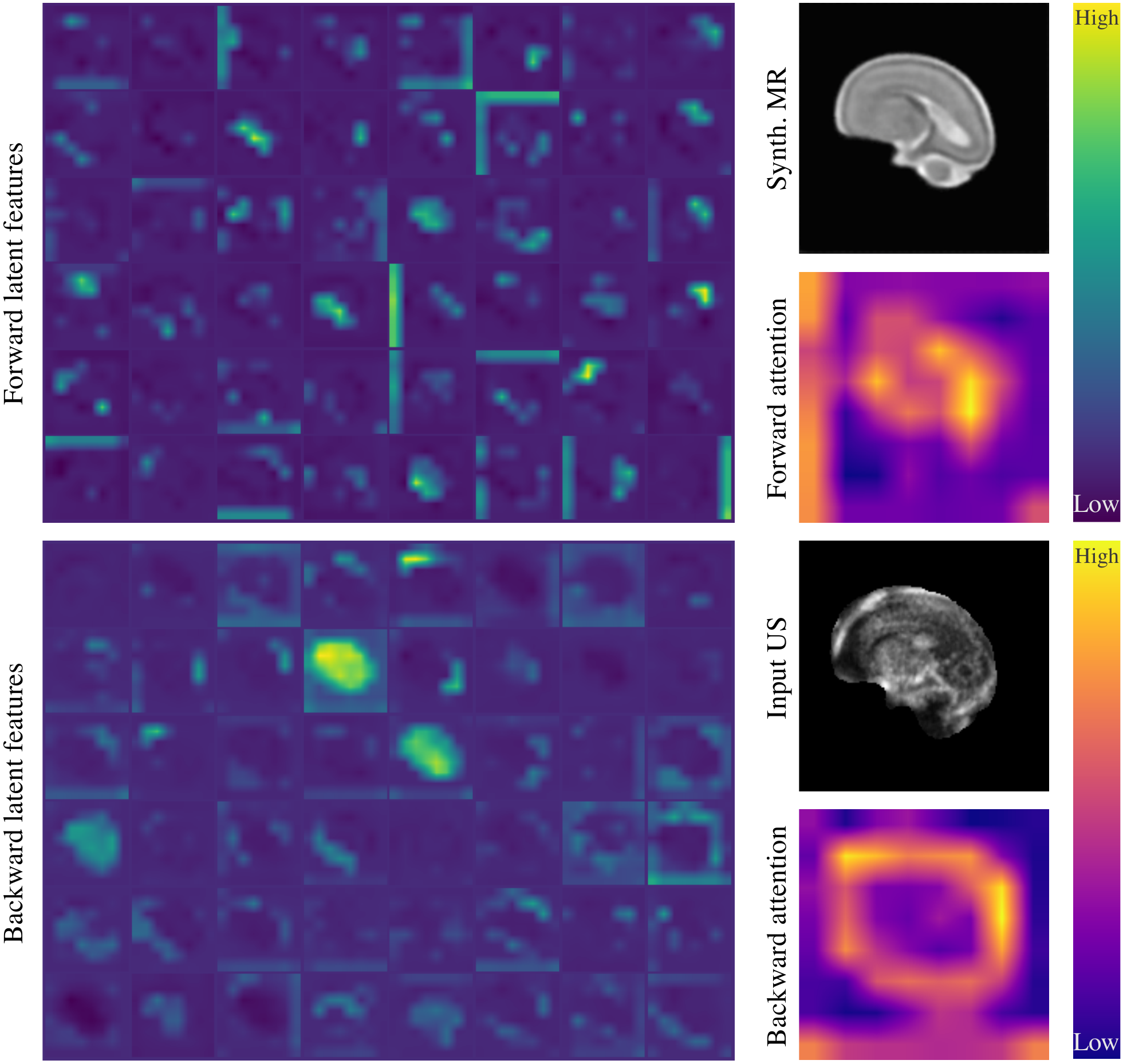}
  \caption{Anatomical latent space visualisation. The top part shows the features (a grid of 48 feature maps) for the forward pass, i.e., at the end of Encoder-A, while the lower part shows the features for the backward pass, i.e., at the end of Encoder-D. The corresponding attention maps with the US and MR images are also shown on the side.}
  \label{fig:anatom_vis}
\end{figure}

Our basic assumption is that the US and MR data share a similar anatomical latent space (as illustrated in Fig.~\ref{fig:pipe}). To better understand the anatomical property of the shared latent space, here we explore it by analysing the features between the encoder and decoder. Specifically, we visualise the feature maps at the end of Encoder-A and Encoder-D in Fig.~\ref{fig:anatom_vis}. From the feature visualisation we can see that for the forward pass, the features focus more on the inner-part anatomical structures, while for the backward pass, the features primarily learn the overall structure of the brain. This is mainly due to the forward pass not only needing reconstruct the US itself but also it has to synthesise the corresponding MR image. As a result, the shared anatomical features are learned at this point. On the other hand, the backward pass aims to infer the original US image and is simultaneously constrained by the latent space from the forward pass, thus focusing more on the global structure. In addition to the features in the shared latent space, we also visualise the corresponding attention map (by the approach in~\cite{zagoruyko2016paying}), which represents where the model pays most attention. Similarly, it can be observed that the forward pass focuses more on the internal structure while the backward pass attention depicts the boundary of the brain. The above analysis {provides some evidence to} validate our assumed anatomical latent space. Note that the assumption of bi-directional latent space consistency aims not to force the forward and backward features of the latent space to be \emph{identical}, but to be similar in distribution, as determined by the discriminator.

\section{Experiments}\label{sec:exp}

\begin{figure*}
  \centering
  \includegraphics[width=0.93\textwidth]{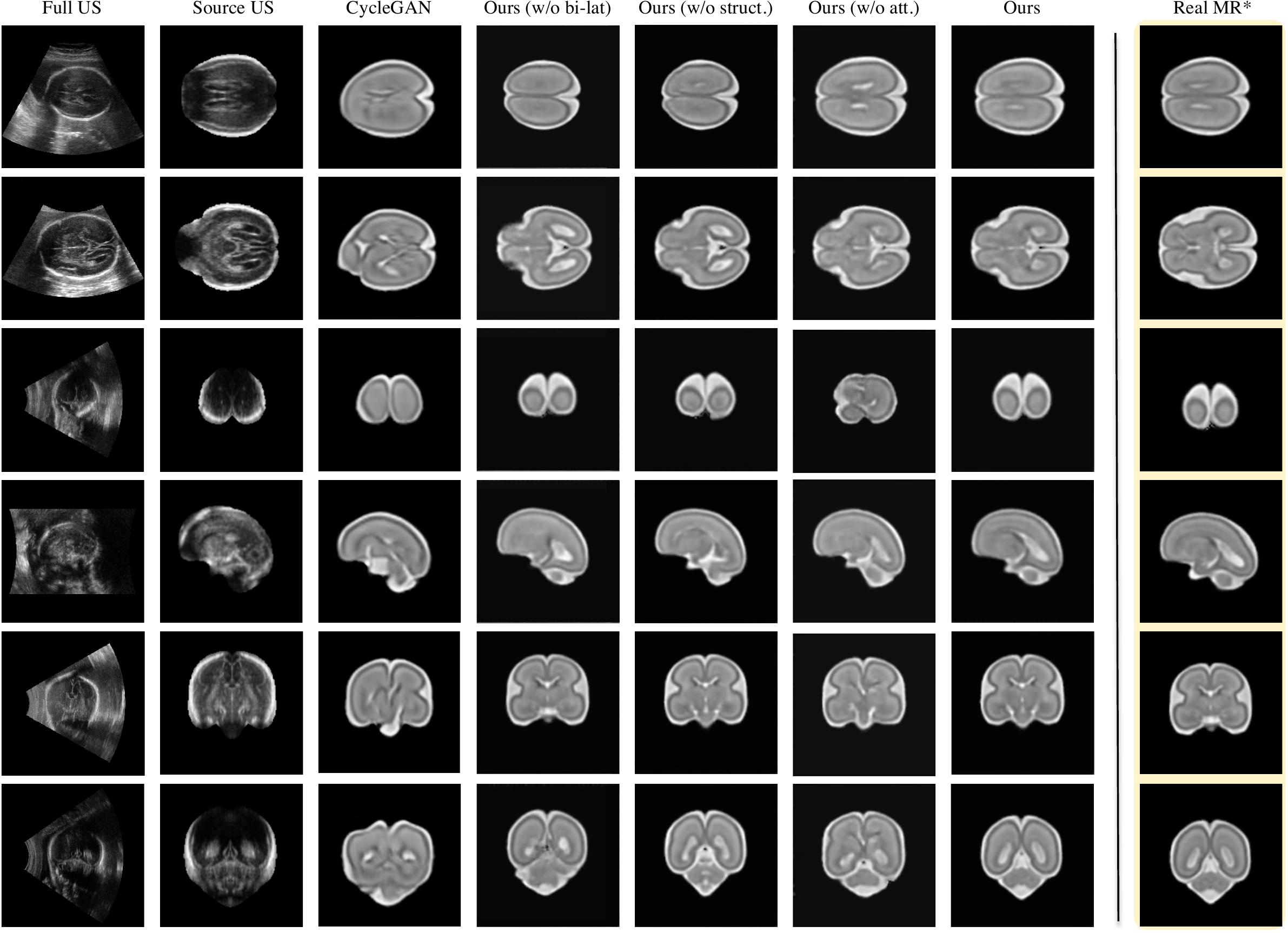}
  \caption{Qualitative performance on the US-to-MR image synthesis. Each row shows an example sample and from left to right: the original full US, pre-processed source US, synthesised MR by CycleGAN~\cite{CycleGAN2017} and our approach (with its counterparts). Some real MR samples are also shown for reference in the last column. *Note that the highlighted \emph{Real~MR} is \textbf{NOT} the exact corresponding MR to the input US, instead illustrative examples for visual comparison.}
  \label{fig:vis}
\end{figure*}

\subsection{Data and Implementation Details}
The training and evaluation of the proposed US-to-MR synthesis framework are based on a dataset consisting of healthy fetal brain US and MR volumes. We obtained the fetal US data from a multi-centre, ethnically diverse dataset~\cite{papageorghiou2014international} of 3D ultrasound scans collected from normal pregnancies. The MR data is obtained from the CRL fetal brain atlas~\cite{gholipour2014construction} database and additional data scanned at Hammersmith Hospital. As proof of principle, we selected the gestational age of 23-week for US and MR data. In the aggregate, the whole dataset consists of 107 US and 2 MRI volumes. Around 36,000 2D US slices and 600 MRI slices were extracted accordingly. 80\% of the whole database accounted for the training and validation sets, while the remaining 20\% acted as the testing set. As detailed in Table~\ref{tab:net}, the proposed model was implemented by simple \emph{conv, up-conv}, and \emph{maxpool} layers. Skip connections were included for each encoder-decoder pair to enhance the structural details. The balancing weights $\lambda$ and $\gamma$ were empirically set to 10 and 1, respectively. All the images (US, MRI, and the corresponding edge maps) used in our framework are of size $160\times160$ {pixels}. The index range for the index prediction task is 160. The learning rate was initialised as $10^{-4}$ and decayed by half for every 20 training epochs. The whole model was trained for 100 epochs. The generator and discriminator are optimised iteratively, i.e. updating the generator for every update of the discriminator. Our whole model was implemented using the PyTorch framework and trained on an Nvidia Titan V GPU in an end-to-end manner. Taking an US image as input, only the A, B, C and Attention blocks in Fig.~\ref{fig:frame} are remained during the model inference, without the whole backward path and all the discriminators.
Implementation of the proposed approach is available online\footnote{\url{https://bitbucket.org/JianboJiao/ssus2mri}}.

\subsection{Evaluation Metrics}\label{sec:metric}
The commonly used evaluation metrics like PSNR (Peak Signal-to-Noise Ratio) or SSIM (Structural Similarity) are not applicable in our study, as US-MR data is not paired. As a result, we propose to leverage two other alternative metrics for the quality evaluation of our synthesised MR images: 1) the MOS (Mean Opinion Score) and 2) the Deformation score (registration metric based on Jacobian).
The MOS measures the quality of a given image by a rating score between 1 and 5: 1 indicates \emph{inferior} while 5 indicates \emph{superior}. A user-study was performed to achieve the MOS performance, given participants from two groups (2 medical experts and 11 beginners), in which each observer was shown with 80 samples. For the Deformation score, an FFD-based~\cite{rueckert1999nonrigid} deformable registration was applied to the synthesised MR to register to a real MR at a similar imaging plane. The average Jacobian (normalised to [0,1]) of the required deformation to complete such registration was computed as the score consequently. The underlying assumption is that a synthesised MRI with high-quality tends to have a lower Jacobian when registering to the real MRI.

\begin{table*}
  \caption{Quantitative performance for MRI synthesis with comparison to several alternative approaches on MOS score and deformation score. The standard deviation ($\pm$std.) is also shown. MOS the higher the better, while deformation the lower the better.}
  \label{tab:mos}
  \centering
  \begin{tabular}{ll|ccc|ccc|cc}
    \toprule
    \multicolumn{2}{l|}{Settings} & AE & GAN & CycleGAN & Ours (w/o bi-lat) & Ours (w/o struct.) & Ours (w/o att.) & Ours & Real\\
    \midrule
    & Expert & $1.00\pm0.00$ & $2.05\pm1.12$ & $2.50\pm0.53$ & $3.05\pm0.80$ & $3.45\pm1.30$ & $3.30\pm1.14$ & \textbf{$3.90\pm0.81$} & \textcolor{gray}{$4.35\pm0.97$}\\
    \multirow{-2.25}{*}{\begin{sideways} MOS$\uparrow$\end{sideways}} & Beginner & $1.01\pm0.03$ & $2.75\pm1.28$ & $3.42\pm0.36$ & $3.69\pm0.71$ & $3.87\pm0.74$ & $3.65\pm0.73$ & \textbf{$4.08\pm0.42$} & \textcolor{gray}{$4.23\pm0.67$}\\
    \midrule
    \multicolumn{2}{l|}{Deformation~$\downarrow$} & $0.97\pm0.09$ & $0.78\pm0.46$ & $0.66\pm0.46$ & $0.55\pm0.22$ & $0.65\pm0.39$ & $0.47\pm0.31$ & \textbf{$0.46\pm0.24$} & \textcolor{gray}{$0.00\pm0.00$} \\
    \bottomrule
  \end{tabular}
\end{table*}

\subsection{Qualitative and Quantitative Performance}
In this section, we firstly present qualitative results of the synthesised MR and secondly quantitatively evaluate synthesis results. In the testing phase, given a test 2D US image as input, the corresponding MR image is synthesised accordingly. Several US example inputs with corresponding synthesised MR images are shown in Fig.~\ref{fig:vis}. Note that the last column (\emph{Real MR}) is \textbf{not} in direct correspondence to the input US, instead is only presented for reference.
These reference MR images are selected from a similar 3D position to the input US images. As a result, although these \emph{Real MR} images are not perfectly aligned to the input US, we assume they are valid references for readers to compare the visual performance.
It can be observed from the results in Fig.~\ref{fig:vis} that the visual appearance of our synthesised MR images is very similar to the real ones. Further the results generated using our approach are visually superior to the results from an alternative approach, CycleGAN~\cite{CycleGAN2017} that is widely used for image synthesis tasks. Additionally, the anatomical structures between the source US and the synthetic MR are well preserved.

The quantitative performance is reported in Table~\ref{tab:mos}, where the evaluation metrics of MOS and deformation are presented. Furthermore, the proposed approach is compared with several alternative methods including an autoencoder (\emph{AE}), GAN~\cite{goodfellow2014generative}, and CycleGAN~\cite{CycleGAN2017}. The presented results suggest that the performance of the proposed US-to-MR synthesis framework surpasses the other CNN-based architectures.

\subsection{Ablation Study}
To better understand the effectiveness of each proposed components, we performed an ablation study by removing each component at a time: the bi-directional latent consistency module (\emph{w/o bi-lat}), the structural consistency module (\emph{w/o struct.}), and the cross-modal attention module (\emph{w/o att.}). The corresponding qualitative and quantitative results are shown in Fig.~\ref{fig:vis} and Table~\ref{tab:mos}, respectively. It can be observed from the results that the model performs worse when removing any of the above components. Specifically, the bi-directional latent space and the cross-modal attention contribute the most to the model performance, which validates our initial assumption of the importance of the shared anatomical space. The structural consistency contributes more to the detailed structures, which is revealed by the deformation metric.
The above qualitative and quantitative results support the inclusion of each proposed component in our model.

\begin{figure}
  \centering
  \includegraphics[width=0.9\columnwidth]{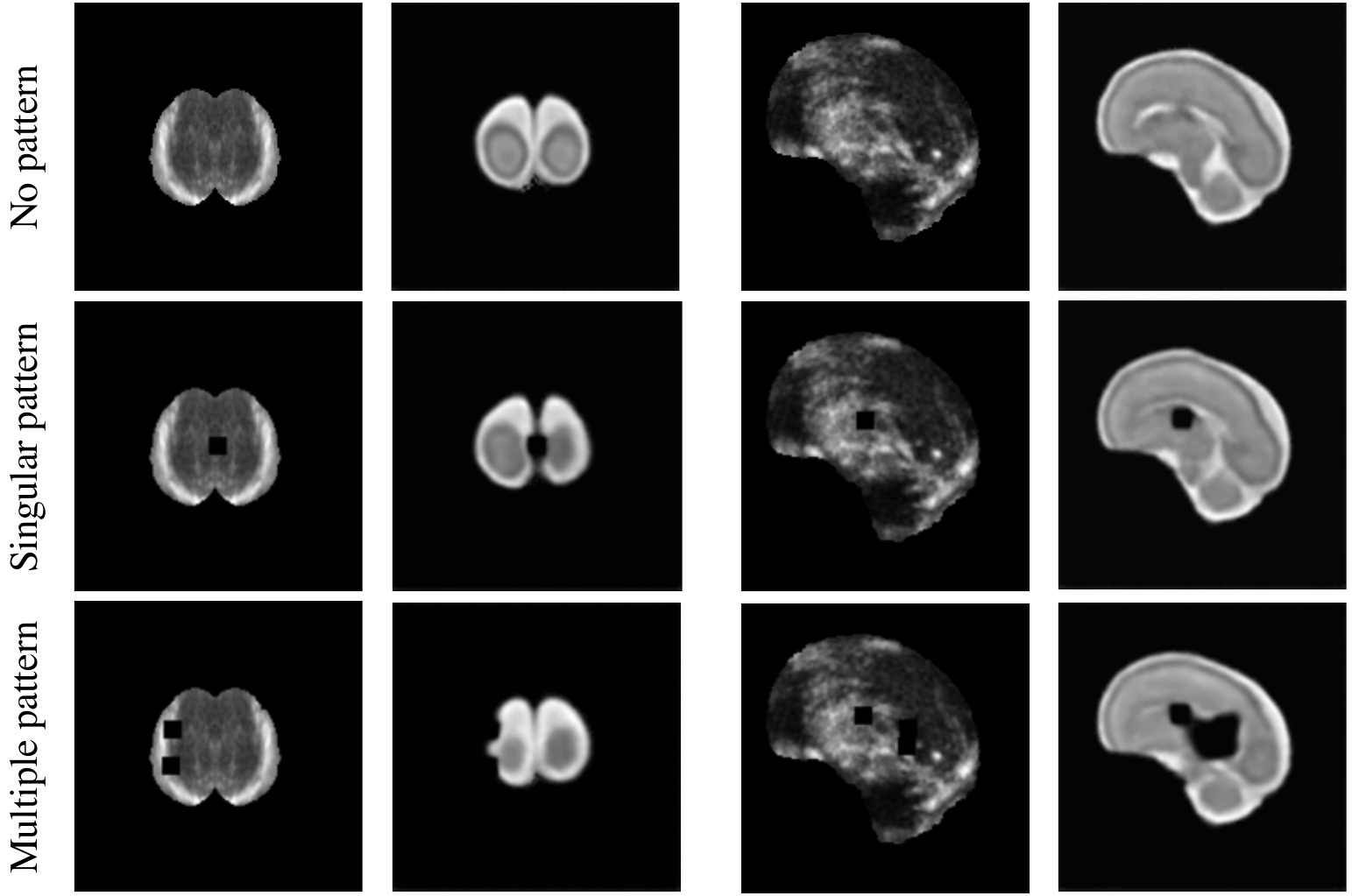}
  \caption{Anatomy-preserving performance for US-to-MR image synthesis on synthetic patterns. For each example the first column shows the input US and the second column the synthesised MRI by our approach.}
  \label{fig:synth}
\end{figure}

\subsection{Anatomy-Preserving Analysis}
In order to evaluate the performance of our approach with respect to the anatomy-preserving property, we perform an analysis based on synthetic abnormal data and real pseudo-paired data:

\paragraph{Synthetic Abnormal Data}
We first randomly apply some synthetic patterns to the input US images and evaluate whether these patterns are preserved during the synthesis of the corresponding MR images.
While various patterns can be applied, here for simplicity, we use a square pattern, which we apply in two ways: singular pattern and multiple patterns.
Note that our trained model is directly applied to these data without any fine-tuning.
Some examples are illustrated in Fig.~\ref{fig:synth}.
We can see from the results that the applied pattern regions are well preserved in the synthesised MR images. Note that in the second example of the multiple-pattern, a larger pattern is generated in the MRI compared to that in the input US. We speculate that this is caused by the small distance between the multiple patterns that makes the synthesis ambiguous for the network.
To quantify the anatomy-preserving property of the proposed model, we further calculate the similarity between the original pattern (multiple version) and the patterns preserved in the synthesised MRI, using the same testing set as aforementioned.
The corresponding quantitative result is reported in Table~\ref{tab:syns_pat}, where we perform a comparison to the alternative architectures of AE, GAN~\cite{goodfellow2014generative}, and CycleGAN~\cite{CycleGAN2017}.
We can see from the result that our approach performs much better than the others, which reveals the anatomy-preserving property of the proposed method.

\begin{table}
  \caption{Quantitative evaluation on our synthesised MR images for synthetic pattern preserving analysis.}
  \label{tab:syns_pat}
  \centering
  \begin{adjustbox}{max width=\columnwidth}
  \begin{tabular}{@{}l|cccc@{}}
    \toprule
    Settings & AE & GAN & CycleGAN & Ours\\
    \midrule
    PSNR (dB)$\uparrow$ & $31.56\pm3.91$ & $34.63\pm13.05$ & $43.07\pm15.07$ & \textbf{$99.37\pm1.55$} \\
    \bottomrule
  \end{tabular}
\end{adjustbox}
\end{table}

\begin{figure}
  \centering
  \includegraphics[width=\columnwidth]{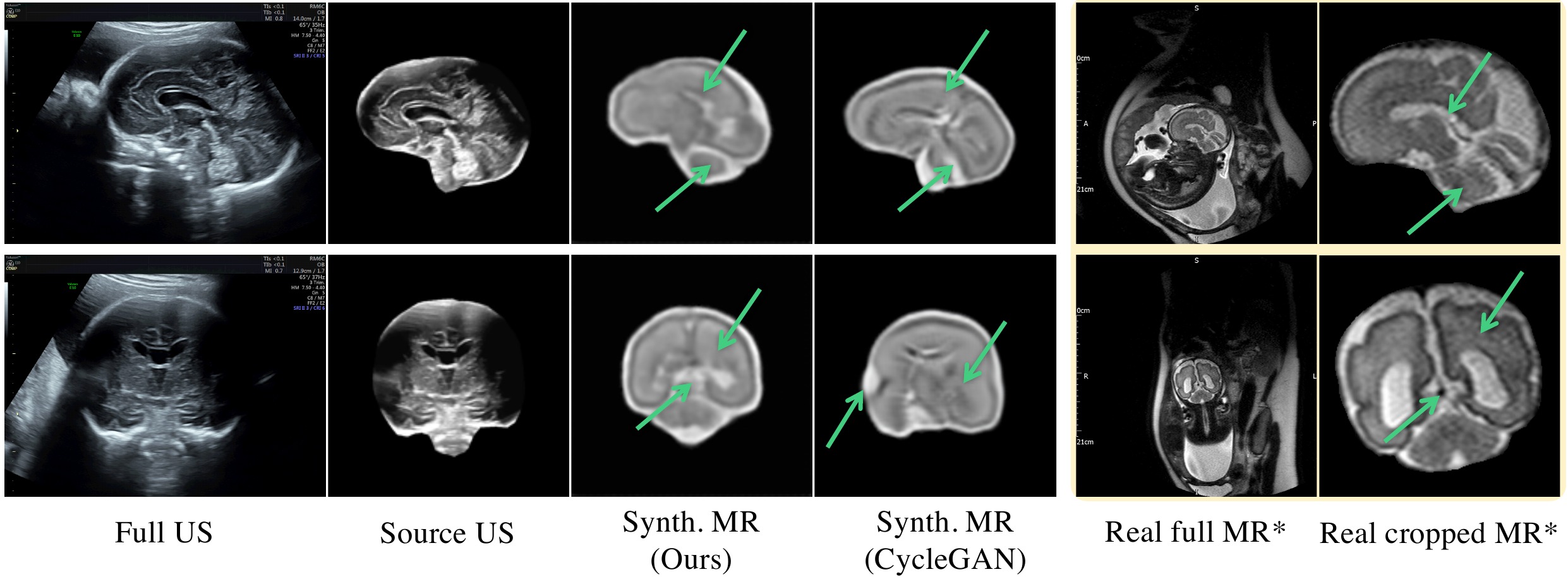}
  \caption{Anatomy-preserving performance for US-to-MR image synthesis on real pseudo-paired data. Key anatomical structures (marked by arrows) are preserved when compared to the real MR images. *The presented MR examples on the right side are not exactly {aligned} to the left US images.}
  \label{fig:real}
\end{figure}

\paragraph{Real Pseudo-Paired Data}
We obtained some anonymised US data with corresponding MRI of the same subject from the John Radcliffe Hospital. Such data can be considered as ``pseudo-paired'', as these US-MR pairs are not exactly corresponding (infeasible to be captured at the exactly same time).
We first pre-process the data by the same scheme as aforementioned and then feed the US data into our network.
Example synthetic MR results are shown in Fig.~\ref{fig:real}, in which we can see that the anatomical structures are well preserved by the synthesis process, with comparison to CycleGAN and the real MR data from the same subject (right side).
Note that our trained model is directly applied to these data without any fine-tuning.
In addition, we also present a quantitative {comparison} to other methods in Table~\ref{tab:ssim}. Specifically, a registration~\cite{aganj2017mid} is performed between the synthesised MR images and the real cropped MR images and the SSIM (Structural Similarity Index Measure) metric is used to measure the performance. We can see that the proposed method outperforms the other alternative solutions, which again validates the effectiveness of our approach.

\begin{table}
  \caption{Quantitative evaluation for US-to-MR image synthesis on real pseudo-paired data.}
  \label{tab:ssim}
  \centering
  \begin{adjustbox}{max width=\columnwidth}
  \begin{tabular}{@{}l|cccc@{}}
    \toprule
    Settings & AE & GAN & CycleGAN & Ours\\
    \midrule
    SSIM~$\uparrow$ & 0.1595$\pm0.0127$ & 0.2271$\pm0.0702$ & 0.6041$\pm0.0483$ & \textbf{0.6250$\pm0.0586$} \\
    \bottomrule
  \end{tabular}
\end{adjustbox}
\end{table}

\subsection{3D Auxiliary Analysis}
\begin{figure}
  \centering
  \includegraphics[width=\columnwidth]{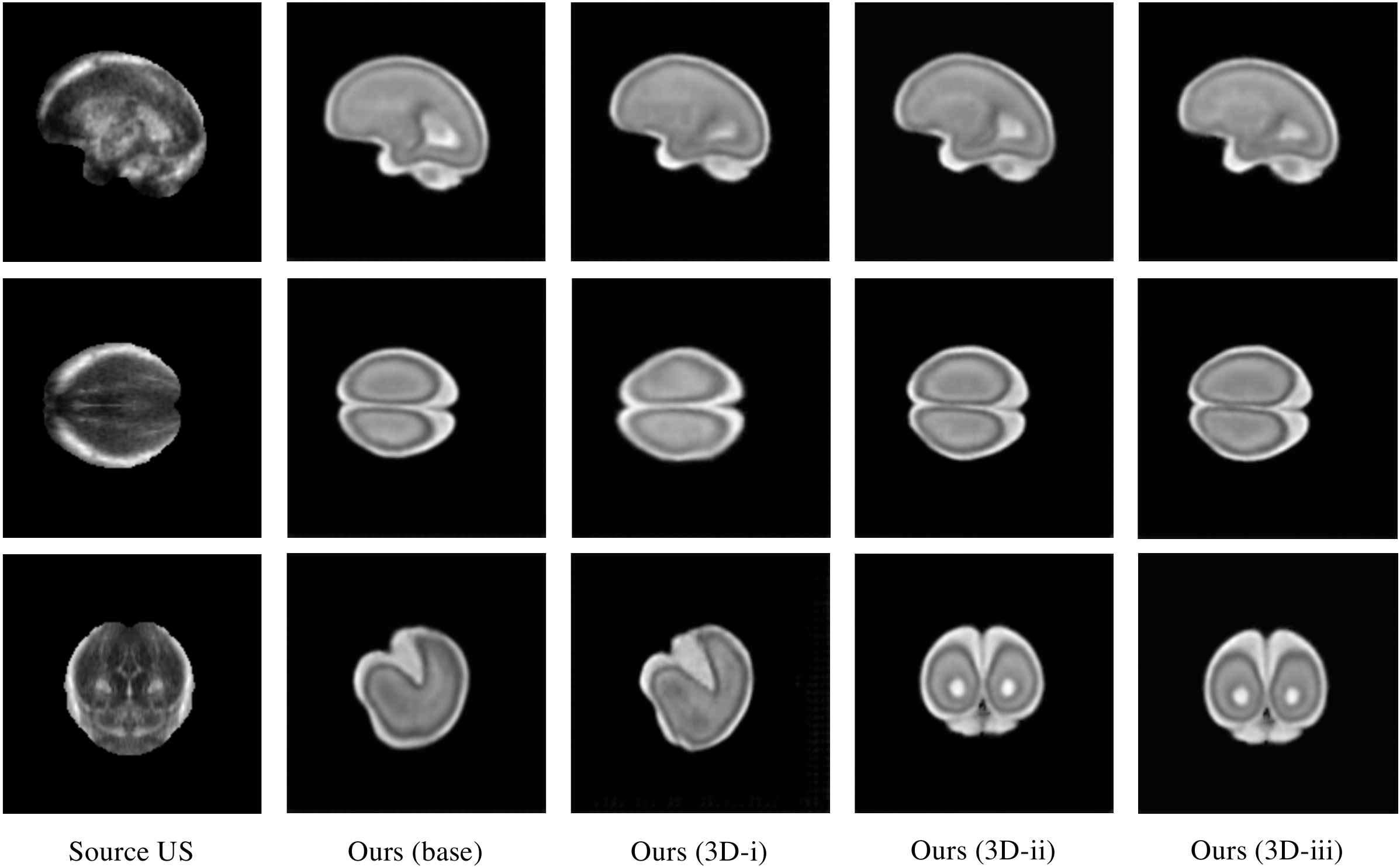}
  \caption{Qualitative performance for US-to-MR image synthesis by 3D auxiliaries. Our base model without 3D auxiliary is included for comparison.}
  \label{fig:3dvis}
\end{figure}

When 3D volumetric US data is available, our synthesis framework can be easily adapted to leverage the additional information, as described in Section~\ref{sec:3D}. Here we analyse the effectiveness of including such auxiliary information. Specifically, we present the deformation score of the three approaches (see Fig.~\ref{fig:frame}-right) with comparison to our 2D-based approach, in Table~\ref{tab:3D}. Qualitative results with one failure case (the third row) of our base model are also shown in Fig.~\ref{fig:3dvis}.
From the results, we can observe that the 3D-ii solution (i.e., predicting the slice index as an auxiliary task) performs the best among all the solutions.
Although both augmenting with neighbouring slices (3D-i) and slice index prediction (3D-ii) provide additional 3D guidance for the target task, by feeding additional slices, ambiguity is also introduced for the discriminative prediction, which leads to slightly worse performance.
On the other hand, the task of directly reasoning the slice index is based on the features from the latent space, which shares anatomical information with the two data modalities and is a less ambiguous task.
Note that all these 3D-auxiliary solutions perform better than the other alternative architectures shown in Table~\ref{tab:mos}.
The slice index reasoning task also performs quite well, with an accuracy of 86\%. Through our experiment, we found that lower performance in this task leads to a less accurate image synthesis task, which on the other side validates the effectiveness of the slice index reasoning.

\begin{table}[t]
  \caption{Quantitative evaluation on our synthesised MR images with comparison to the 3D auxiliary approaches. 3D-i, ii, iii are the corresponding approaches shown in Fig.~\ref{fig:frame} (Right).}
  \label{tab:3D}
  \centering
  \resizebox{\columnwidth}{!}{
  \begin{tabular}{@{}l|cccc@{}}
    \toprule
    Settings & Ours (base) & Ours (3D-i) & Ours (3D-ii) & Ours (3D-iii)\\
    \midrule
    Deformation~$\downarrow$ & $0.46\pm0.24$ & $0.59\pm0.33$ & $0.44\pm0.19$ & $0.60\pm0.37$\\
    \bottomrule
  \end{tabular}
  }
\end{table}

\section{Discussions}\label{sec:app}
\subsection{Potential Applications}
\paragraph{Annotation Transfer between US and MRI}
Data annotation from human experts is a long-standing challenge for data-driven medical image models, especially for those data (e.g., US) where anatomical structures are difficult to recognise.
Thus, well-trained experts are necessary to annotate such data, which is labour-intensive work. Even though, the accuracy of the annotation cannot be always guaranteed.
On the other hand, if the non-anatomical data has an anatomical correspondence in which annotation is much easier, the labelling effort can be largely mitigated.
By using the model proposed in this work, corresponding MRI data can be generated for each US image. Since the data annotation for MRI is much more efficient compared to that for US, the annotation work can be first done on MRI and then transferred to the US.

\paragraph{Data Augmentation for Learning-Based Approaches}
A large amount of data is essential for deep learning models, whereas the acquisition of some data modalities is rather challenging.
Fetal MRI is a kind of such data, where routinely MRI scan is usually not provided in most cases.
As a result, data-driven deep learning models for fetal MRI related problems are infeasible to scale.
In this case, if the correlation between fetal MRI and sufficient routine US scans can be bridged, training data for MRI can be generated.
The proposed approach in this work may be well-suited for this. By feeding US images into our model, abundant corresponding MRI-like data can be synthesised to train deep learning models for fetal MRI.

\subsection{Generalisation}
The input to our model is pre-processed (cropped and aligned) US data instead of full US scans. Although promising synthesis results have been achieved, one limitation of our model lies in the input data assumptions. The generalisation to full US scans will require further research.
Since the proposed method is data-driven and our model was trained with 23-week data, it is not an optimal model for other gestational ages. Similarly, as our model is trained on healthy data, it may not be the optimal model for new US images with real pathologies. This is a challenging problem for modern deep learning approaches and could be possibly addressed by transfer learning strategies.
Without real paired data for evaluation, we cannot determine if the synthesised images are realistic enough to transfer all useful diagnostic information. Looking into  this would be a natural next step towards assessing potential clinical utility.
Finally, the proposed approach is a general framework that can be readily extended to other body parts (e.g., heart), medical imaging modalities (e.g., CT) and clinical application domains.

\section{Conclusion}\label{sec:conc}
In this paper, we have presented an original method to synthesise MR-like fetal brain images from unpaired US images, via a novel anatomy-aware self-supervised framework. Specifically, shared latent features between the two modalities (US and MR) are first extracted, from which the target MRI is synthesised under a group of anatomy-aware constraints. A cross-modal attention module is introduced to incorporate non-local guidance across the two modalities. An investigation to leverage 3D volumetric auxiliaries is also presented. Experimental results demonstrate the effectiveness of the proposed framework both qualitatively and quantitatively, with comparison to alternative CNN architectures.

We believe the proposed framework to be useful within analysis tasks such as the alignment between US and MRI and for communicating US findings to obstetricians and patients. The generalisation to full US images is another interesting direction worth investigation. Given more paired examples in the future, the synthesis quality would also be improved.

% use section* for acknowledgment
\section*{Acknowledgment}
The authors would like to thank Andrew Zisserman for many helpful discussions, the volunteers for assessing images, and NVIDIA Corporation for the Titan V GPU donation.
Ana Namburete is grateful for support from the UK Royal Academy of Engineering under its Engineering for Development Research Fellowships scheme.

% references section

% can use a bibliography generated by BibTeX as a .bbl file
% BibTeX documentation can be easily obtained at:
% http://mirror.ctan.org/biblio/bibtex/contrib/doc/
% The IEEEtran BibTeX style support page is at:
% http://www.michaelshell.org/tex/ieeetran/bibtex/
\bibliographystyle{IEEEtran}
% argument is your BibTeX string definitions and bibliography database(s)
\bibliography{mybib}

% Generated by IEEEtran.bst, version: 1.14 (2015/08/26)
\begin{thebibliography}{10}
\providecommand{\url}[1]{#1}
\csname url@samestyle\endcsname
\providecommand{\newblock}{\relax}
\providecommand{\bibinfo}[2]{#2}
\providecommand{\BIBentrySTDinterwordspacing}{\spaceskip=0pt\relax}
\providecommand{\BIBentryALTinterwordstretchfactor}{4}
\providecommand{\BIBentryALTinterwordspacing}{\spaceskip=\fontdimen2\font plus
\BIBentryALTinterwordstretchfactor\fontdimen3\font minus
  \fontdimen4\font\relax}
\providecommand{\BIBforeignlanguage}[2]{{%
\expandafter\ifx\csname l@#1\endcsname\relax
\typeout{** WARNING: IEEEtran.bst: No hyphenation pattern has been}%
\typeout{** loaded for the language `#1'. Using the pattern for}%
\typeout{** the default language instead.}%
\else
\language=\csname l@#1\endcsname
\fi
#2}}
\providecommand{\BIBdecl}{\relax}
\BIBdecl

\bibitem{pugash2008prenatal}
D.~Pugash, P.~C. Brugger, D.~Bettelheim, and D.~Prayer, ``Prenatal ultrasound
  and fetal mri: the comparative value of each modality in prenatal
  diagnosis,'' \emph{European journal of radiology}, vol.~68, no.~2, pp.
  214--226, 2008.

\bibitem{bulas2013benefits}
D.~Bulas and A.~Egloff, ``Benefits and risks of mri in pregnancy,'' in
  \emph{Seminars in perinatology}, vol.~37, no.~5.\hskip 1em plus 0.5em minus
  0.4em\relax Elsevier, 2013, pp. 301--304.

\bibitem{zhao2018towards}
Y.~Zhao \emph{et~al.}, ``Towards mr-only radiotherapy treatment planning:
  Synthetic ct generation using multi-view deep convolutional neural
  networks,'' in \emph{MICCAI}, 2018, pp. 286--294.

\bibitem{nie2017medical}
D.~Nie \emph{et~al.}, ``Medical image synthesis with context-aware generative
  adversarial networks,'' in \emph{MICCAI}, 2017, pp. 417--425.

\bibitem{yang2018unpaired}
H.~Yang \emph{et~al.}, ``Unpaired brain mr-to-ct synthesis using a
  structure-constrained cyclegan,'' in \emph{Deep Learning in Medical Image
  Analysis and Multimodal Learning for Clinical Decision Support}, 2018, pp.
  174--182.

\bibitem{costa2018end}
P.~Costa \emph{et~al.}, ``End-to-end adversarial retinal image synthesis,''
  \emph{IEEE TMI}, vol.~37, no.~3, pp. 781--791, 2018.

\bibitem{costa2017towards}
------, ``Towards adversarial retinal image synthesis,'' \emph{arXiv preprint
  arXiv:1701.08974}, 2017.

\bibitem{kuklisova2013registration}
M.~Kuklisova-Murgasova \emph{et~al.}, ``Registration of 3d fetal
  neurosonography and mri,'' \emph{Medical image analysis}, vol.~17, no.~8, pp.
  1137--1150, 2013.

\bibitem{king2010registering}
A.~P. King \emph{et~al.}, ``Registering preprocedure volumetric images with
  intraprocedure 3-d ultrasound using an ultrasound imaging model,'' \emph{IEEE
  TMI}, vol.~29, no.~3, pp. 924--937, 2010.

\bibitem{berker2012mri}
Y.~Berker \emph{et~al.}, ``Mri-based attenuation correction for hybrid pet/mri
  systems: a 4-class tissue segmentation technique using a combined
  ultrashort-echo-time/dixon mri sequence,'' \emph{Journal of nuclear
  medicine}, vol.~53, no.~5, p. 796, 2012.

\bibitem{delpon2016comparison}
G.~Delpon \emph{et~al.}, ``Comparison of automated atlas-based segmentation
  software for postoperative prostate cancer radiotherapy,'' \emph{Frontiers in
  oncology}, vol.~6, p. 178, 2016.

\bibitem{sjolund2015generating}
J.~Sj{\"o}lund, D.~Forsberg, M.~Andersson, and H.~Knutsson, ``Generating
  patient specific pseudo-ct of the head from mr using atlas-based
  regression,'' \emph{Physics in Medicine \& Biology}, vol.~60, no.~2, p. 825,
  2015.

\bibitem{catana2010towards}
C.~Catana \emph{et~al.}, ``Towards implementing an mr-based pet attenuation
  correction method for neurological studies on the mr-pet brain prototype,''
  \emph{Journal of nuclear medicine}, vol.~51, no.~9, p. 1431, 2010.

\bibitem{roy2017synthesizing}
S.~Roy, J.~A. Butman, and D.~L. Pham, ``Synthesizing ct from ultrashort
  echo-time mr images via convolutional neural networks,'' in
  \emph{International Workshop on Simulation and Synthesis in Medical Imaging},
  2017, pp. 24--32.

\bibitem{zhang2018translating}
Z.~Zhang, L.~Yang, and Y.~Zheng, ``Translating and segmenting multimodal
  medical volumes with cycle-and shape-consistency generative adversarial
  network,'' in \emph{CVPR}, 2018, pp. 9242--9251.

\bibitem{CycleGAN2017}
J.-Y. Zhu, T.~Park, P.~Isola, and A.~A. Efros, ``Unpaired image-to-image
  translation using cycle-consistent adversarial networkss,'' in \emph{ICCV},
  2017.

\bibitem{Jianbo2019anatomy}
J.~Jiao, A.~I. Namburete, A.~T. Papageorghiou, and J.~A. Noble, ``Anatomy-aware
  self-supervised fetal mri synthesis from unpaired ultrasound images,'' in
  \emph{International Workshop on Machine Learning in Medical Imaging}.\hskip
  1em plus 0.5em minus 0.4em\relax Springer, 2019.

\bibitem{lee2003radiotherapy}
Y.~K. Lee \emph{et~al.}, ``Radiotherapy treatment planning of prostate cancer
  using magnetic resonance imaging alone,'' \emph{Radiotherapy and oncology},
  vol.~66, no.~2, pp. 203--216, 2003.

\bibitem{dowling2012atlas}
J.~A. Dowling \emph{et~al.}, ``An atlas-based electron density mapping method
  for magnetic resonance imaging (mri)-alone treatment planning and adaptive
  mri-based prostate radiation therapy,'' \emph{International Journal of
  Radiation Oncology* Biology* Physics}, vol.~83, no.~1, pp. e5--e11, 2012.

\bibitem{goodfellow2014generative}
I.~Goodfellow \emph{et~al.}, ``Generative adversarial nets,'' in
  \emph{NeurIPS}, 2014, pp. 2672--2680.

\bibitem{bengio2007greedy}
Y.~Bengio, P.~Lamblin, D.~Popovici, and H.~Larochelle, ``Greedy layer-wise
  training of deep networks,'' in \emph{NeurIPS}, 2007, pp. 153--160.

\bibitem{vincent2010stacked}
P.~Vincent, H.~Larochelle, I.~Lajoie, Y.~Bengio, and P.-A. Manzagol, ``Stacked
  denoising autoencoders: Learning useful representations in a deep network
  with a local denoising criterion,'' \emph{Journal of machine learning
  research}, vol.~11, no. Dec, pp. 3371--3408, 2010.

\bibitem{kingma2013auto}
D.~P. Kingma and M.~Welling, ``Auto-encoding variational bayes,'' \emph{arXiv
  preprint arXiv:1312.6114}, 2013.

\bibitem{yi2017dualgan}
Z.~Yi, H.~Zhang, P.~Tan, and M.~Gong, ``Dualgan: Unsupervised dual learning for
  image-to-image translation,'' in \emph{ICCV}, 2017, pp. 2849--2857.

\bibitem{wolterink2017deep}
J.~M. Wolterink, A.~M. Dinkla, M.~H. Savenije, P.~R. Seevinck, C.~A. van~den
  Berg, and I.~I{\v{s}}gum, ``Deep mr to ct synthesis using unpaired data,'' in
  \emph{International Workshop on Simulation and Synthesis in Medical
  Imaging}.\hskip 1em plus 0.5em minus 0.4em\relax Springer, 2017, pp. 14--23.

\bibitem{hiasa2018cross}
Y.~Hiasa, Y.~Otake, M.~Takao, T.~Matsuoka, K.~Takashima, A.~Carass, J.~L.
  Prince, N.~Sugano, and Y.~Sato, ``Cross-modality image synthesis from
  unpaired data using cyclegan,'' in \emph{International workshop on simulation
  and synthesis in medical imaging}.\hskip 1em plus 0.5em minus 0.4em\relax
  Springer, 2018, pp. 31--41.

\bibitem{chartsias2017adversarial}
A.~Chartsias, T.~Joyce, R.~Dharmakumar, and S.~A. Tsaftaris, ``Adversarial
  image synthesis for unpaired multi-modal cardiac data,'' in
  \emph{International workshop on simulation and synthesis in medical
  imaging}.\hskip 1em plus 0.5em minus 0.4em\relax Springer, 2017, pp. 3--13.

\bibitem{zeng2019hybrid}
G.~Zeng and G.~Zheng, ``Hybrid generative adversarial networks for deep mr to
  ct synthesis using unpaired data,'' in \emph{International Conference on
  Medical Image Computing and Computer-Assisted Intervention}.\hskip 1em plus
  0.5em minus 0.4em\relax Springer, 2019, pp. 759--767.

\bibitem{maraci2014searching}
M.~A. Maraci, R.~Napolitano, A.~Papageorghiou, and J.~A. Noble, ``Searching for
  structures of interest in an ultrasound video sequence,'' in
  \emph{International Workshop on Machine Learning in Medical Imaging}.\hskip
  1em plus 0.5em minus 0.4em\relax Springer, 2014, pp. 133--140.

\bibitem{yaqub2015guided}
M.~Yaqub, B.~Kelly, A.~T. Papageorghiou, and J.~A. Noble, ``Guided random
  forests for identification of key fetal anatomy and image categorization in
  ultrasound scans,'' in \emph{MICCAI}.\hskip 1em plus 0.5em minus 0.4em\relax
  Springer, 2015, pp. 687--694.

\bibitem{chen2015standard}
H.~Chen \emph{et~al.}, ``Standard plane localization in fetal ultrasound via
  domain transferred deep neural networks,'' \emph{IEEE journal of biomedical
  and health informatics}, vol.~19, no.~5, pp. 1627--1636, 2015.

\bibitem{baumgartner2017sononet}
C.~F. Baumgartner \emph{et~al.}, ``Sononet: real-time detection and
  localisation of fetal standard scan planes in freehand ultrasound,''
  \emph{IEEE TMI}, vol.~36, no.~11, pp. 2204--2215, 2017.

\bibitem{cai2018multi}
Y.~Cai, H.~Sharma, P.~Chatelain, and J.~A. Noble, ``Multi-task sonoeyenet:
  detection of fetal standardized planes assisted by generated sonographer
  attention maps,'' in \emph{MICCAI}.\hskip 1em plus 0.5em minus 0.4em\relax
  Springer, 2018, pp. 871--879.

\bibitem{wein2008automatic}
W.~Wein, S.~Brunke, A.~Khamene, M.~R. Callstrom, and N.~Navab, ``Automatic
  ct-ultrasound registration for diagnostic imaging and image-guided
  intervention,'' \emph{Medical image analysis}, vol.~12, no.~5, pp. 577--585,
  2008.

\bibitem{xiao2019evaluation}
Y.~Xiao \emph{et~al.}, ``Evaluation of mri to ultrasound registration methods
  for brain shift correction: The curious2018 challenge,'' \emph{arXiv preprint
  arXiv:1904.10535}, 2019.

\bibitem{namburete2018fully}
A.~I. Namburete, W.~Xie, M.~Yaqub, A.~Zisserman, and J.~A. Noble,
  ``Fully-automated alignment of 3d fetal brain ultrasound to a canonical
  reference space using multi-task learning,'' \emph{Medical image analysis},
  vol.~46, pp. 1--14, 2018.

\bibitem{zhang2018self}
H.~Zhang, I.~Goodfellow, D.~Metaxas, and A.~Odena, ``Self-attention generative
  adversarial networks,'' \emph{arXiv preprint arXiv:1805.08318}, 2018.

\bibitem{wang2018non}
X.~Wang, R.~Girshick, A.~Gupta, and K.~He, ``Non-local neural networks,'' in
  \emph{CVPR}, 2018, pp. 7794--7803.

\bibitem{zagoruyko2016paying}
S.~Zagoruyko and N.~Komodakis, ``Paying more attention to attention: Improving
  the performance of convolutional neural networks via attention transfer,''
  \emph{arXiv preprint arXiv:1612.03928}, 2016.

\bibitem{papageorghiou2014international}
A.~T. Papageorghiou \emph{et~al.}, ``International standards for fetal growth
  based on serial ultrasound measurements: the fetal growth longitudinal study
  of the intergrowth-21st project,'' \emph{The Lancet}, vol. 384, no. 9946, pp.
  869--879, 2014.

\bibitem{gholipour2014construction}
A.~Gholipour \emph{et~al.}, ``Construction of a deformable spatiotemporal mri
  atlas of the fetal brain: evaluation of similarity metrics and deformation
  models,'' in \emph{MICCAI}.\hskip 1em plus 0.5em minus 0.4em\relax Springer,
  2014, pp. 292--299.

\bibitem{rueckert1999nonrigid}
D.~Rueckert, L.~I. Sonoda, C.~Hayes, D.~L. Hill, M.~O. Leach, and D.~J. Hawkes,
  ``Nonrigid registration using free-form deformations: application to breast
  mr images,'' \emph{IEEE TMI}, vol.~18, no.~8, pp. 712--721, 1999.

\bibitem{aganj2017mid}
I.~Aganj, J.~E. Iglesias, M.~Reuter, M.~R. Sabuncu, and B.~Fischl,
  ``Mid-space-independent deformable image registration,'' \emph{NeuroImage},
  vol. 152, pp. 158--170, 2017.

\end{thebibliography}

\end{document}